\documentclass[conference]{IEEEtran}
\IEEEoverridecommandlockouts
\usepackage{cite}
\usepackage{amsmath,amssymb,amsfonts}
\usepackage{algorithm}
\usepackage{algorithmic}
\usepackage{graphicx}
\usepackage{textcomp}
\usepackage{xcolor}
\usepackage{soul}
\usepackage{enumitem}
\usepackage{pifont}

\usepackage{stfloats}
\usepackage{url}
\usepackage[bottom]{footmisc}

\usepackage{flushend} 

\usepackage{booktabs}
\usepackage{tabularx}
\usepackage{colortbl}

\usepackage[caption=false]{subfig}
\usepackage[font=footnotesize,labelfont=bf]{caption}

\def\BibTeX{{\rm B\kern-.05em{\sc i\kern-.025em b}\kern-.08em
    T\kern-.1667em\lower.7ex\hbox{E}\kern-.125emX}}

\usepackage[absolute]{textpos}

\usepackage{titlesec}

\titlespacing*{\section}{0pt}{4pt}{2pt}
\titlespacing*{\subsection}{0pt}{3pt}{1pt}
\titlespacing*{\subsubsection}{0pt}{2pt}{1pt}

\begin{document}
\begin{textblock}{5}(11.8,0.55)
(Special Session)
\end{textblock}

\begin{textblock}{14}(5.7,0.75)
This paper will be presented at IEEE VLSI Test Symposium (VTS) 2026.
\end{textblock} 


\title{From Language to Logic: Bridging LLMs \& Formal Representations for RTL Assertion Generation}

\author{\IEEEauthorblockN{Nowfel Mashnoor, Hadi Kamali, Kimia Azar}
\IEEEauthorblockA{\textit{Department of Electrical and Computer Engineering (ECE), University of Central Florida, Orlando, FL 32816, USA} \\
\{nowfel, kamali, azar\}@ucf.edu}
}

\maketitle

\begin{abstract}
SystemVerilog Assertions (SVA) are essential for formal verification of digital hardware, yet their manual creation demands significant expertise in both the design under verification and temporal logic. Recent studies have explored using large language models (LLMs) to automate SVA generation, but existing approaches suffer from incorrect signal references, missing timing constraints, and lack of formal correctness guarantees. This paper presents \texttt{ProofLoop}, a tool-augmented \textit{ReAct} agent that generates SVA from natural-language specifications using a \textit{solver-in-the-loop} approach. The agent operates in two phases: \textit{Phase~A} autonomously gathers design context by invoking EDA and formal tools, including semantic search over an AST-indexed vector database and JasperGold structural queries, while \textit{Phase~B} generates SVA and iteratively refines it using JasperGold formal proof feedback over up to fixed (here 3) verification rounds. We evaluate \texttt{ProofLoop} on FVEval Design2SVA design benchmarks and demonstrate that this framework can achieve 93.7\% syntax correctness and 82.0\% functional correctness. An ablation study confirms that each component, i.e., retrieval-augmented generation (RAG), JasperGold tools, and the verification loop contributes significantly (and orthogonally).
\end{abstract}

\begin{IEEEkeywords}
SystemVerilog Assertions, Formal Verification, LLM, ReAct Agent, Solver-in-the-Loop, RAG.
\end{IEEEkeywords}

\section{Introduction}

Assertion-based verification (ABV) is a cornerstone of modern hardware verification, enabling engineers to express design intent as SystemVerilog Assertions (SVA) that can be formally proven or monitored during simulation \cite{witharana2022survey, vijayaraghavan2005sva, clarke2018model}. However, writing quality SVA remains  computation- and labor-intensive that requires deep understanding of both register-transfer level (RTL) semantics and temporal logic \cite{aftabjahani2021special}. Prior automated approaches based on data mining \cite{vasudevan2010goldmine, germiniani2022harm, danese2017ateam} or template matching \cite{orenes2021autosva, farzana2019soc} have shown promise but remain limited in expressiveness and require design-specific configuration. As hardware complexity continues to grow, particularly at system-on-chip (SoC) scale, the cost of non-automated assertion authoring has become a bottleneck in verification.

Large language models (LLMs) have shown promise in automating SVA generation \cite{kande2024security, orenes2023using, sun2023nl2sva}, building on their success in RTL code generation \cite{thakur2023verigen, akyash2025decortl, liu2024rtlcoder, akyash2025rtl++, zhao2025codev, mashnoor2026meltrtl} and verification \cite{huang2024towards, mashnoor2025llm}. However, directly prompting an LLM with raw RTL and a specification often produces assertions with incorrect signal names, wrong clock/reset assumptions, and mismatched temporal sequences (w.r.t. actual hardware behavior) \cite{pulavarthi2025assertionbench, kang2024domain}. These errors persist because LLMs lack access to the structural metadata that verification engineers rely on, e.g., flip-flop clocking, reset values, signal fanin/fanout relationships, and FSM state encodings. While retrieval-augmented generation (RAG) \cite{lewis2020retrieval} has improved LLM grounding in other domains, and recent work has applied it to RTL code generation \cite{deepv2025}, its potential for assertion generation, particularly when combined with formal-tool structural queries remains largely unexplored.

To address these challenges, we present \texttt{\textbf{ProofLoop}}, a tool-augmented ReAct agent for automated SVA generation that integrates Cadence JasperGold as a \textit{solver-in-the-loop}. This agentic approach allows the model to dynamically decide what information it needs, rather than following a fixed pipeline. The key contributions are:

\begin{enumerate}[leftmargin=*]
    \item[(1)] \textbf{\textit{\underline{Agentic context synthesis.}}} We propose a ReAct-based agent that autonomously gathers verification-relevant context via semantic retrieval over an AST-indexed RTL representation and structural queries to a formal tool.
    \item[(2)] \textbf{\textit{\underline{Solver-in-the-loop refinement.}}} We introduce an iterative verification loop that leverages formal proof feedback to correct generated assertions, enabling targeted repair of syntax errors, signal mismatches, and inconsistencies.
    \item[(3)] \textbf{\textit{\underline{Scalable assertion generation.}}} We demonstrate that tool-augmented context and iterative refinement significantly improve both syntactic and functional correctness, with robustness that scales to complex, multi-module designs.
    \item[(4)] \textbf{\textit{\underline{Comprehensive empirical study.}}} We evaluate on the FVEval benchmark \cite{kang2024fveval} and present ablation analyses that quantify the impact of retrieval, structural analysis, and verification feedback, highlighting the role of tool augmentation in bridging natural language and formal specifications.
\end{enumerate}

\section{Related Work}

Traditional approaches to automated assertion generation include GoldMine \cite{vasudevan2010goldmine}, which uses data mining over simulation traces; HARM \cite{germiniani2022harm}, a hint-based assertion miner; and AutoSVA \cite{orenes2021autosva}, which generates formal testbenches from interface annotations. While effective for specific patterns, these methods do not generalize from natural-language specifications.

Table~\ref{tab:related_comparison} summarizes key differences among recent LLM-based SVA generation approaches. ChIRAAG \cite{mali2024chiraag} uses GPT-4 with simulation log feedback. AssertLLM \cite{fang2024assertllm} introduces a three-phase multi-LLM pipeline, but operates without formal feedback. SANGAM \cite{gupta2025sangam} applies Monte Carlo Tree Self-Refine (MCTSr) with JasperGold syntax checking in the search loop; however, its MCTSr-based search is computationally expensive and does not leverage JG metadata for context enrichment. Saarthi \cite{kumar2025saarthi} proposes an agentic multi-agent system with formal proof analysis but lacks a large-scale benchmark evaluation. AssertionForge \cite{bai2025assertionforge} constructs a knowledge graph for multi-resolution context synthesis but does not include iterative formal refinement. SVAgent \cite{guo2025svagent} uses fine-grained prompting for security-oriented assertions without RAG or iterative formal refinement. ChatSVA \cite{chatsva2026} combines task-specific fine-tuned LLMs with data augmentation, complementing our tool-augmented approach. Menon et al.~\cite{menon2025vert} contribute a curated SVA dataset that supports fine-tuning-based methods. FVEval \cite{kang2024fveval} establishes the benchmark and evaluation methodology we adopt. RTLFixer \cite{tsai2023rtlfixer} and Blocklove et al.~\cite{blocklove2025improving} demonstrate that iterative EDA tool feedback can correct LLM-generated RTL, a concept we extend to assertion generation. Hassan et al.~\cite{hassan2024llmguided} combine LLM-guided formal verification with mutation testing, highlighting the synergy between LLMs and formal tools. As shown in Table \ref{tab:related_comparison}, \texttt{\textbf{ProofLoop}} differs from prior work in two key ways: (i)~the agent autonomously decides which tools to invoke through a reasoning loop, dynamically adapting its information-gathering strategy per specification; and (ii)~JasperGold is integrated both for structural queries during context gathering (signal dependency and register analysis) and for formal proof during iterative verification.

\begin{figure*}[b]
\centering
\includegraphics[width=\textwidth]{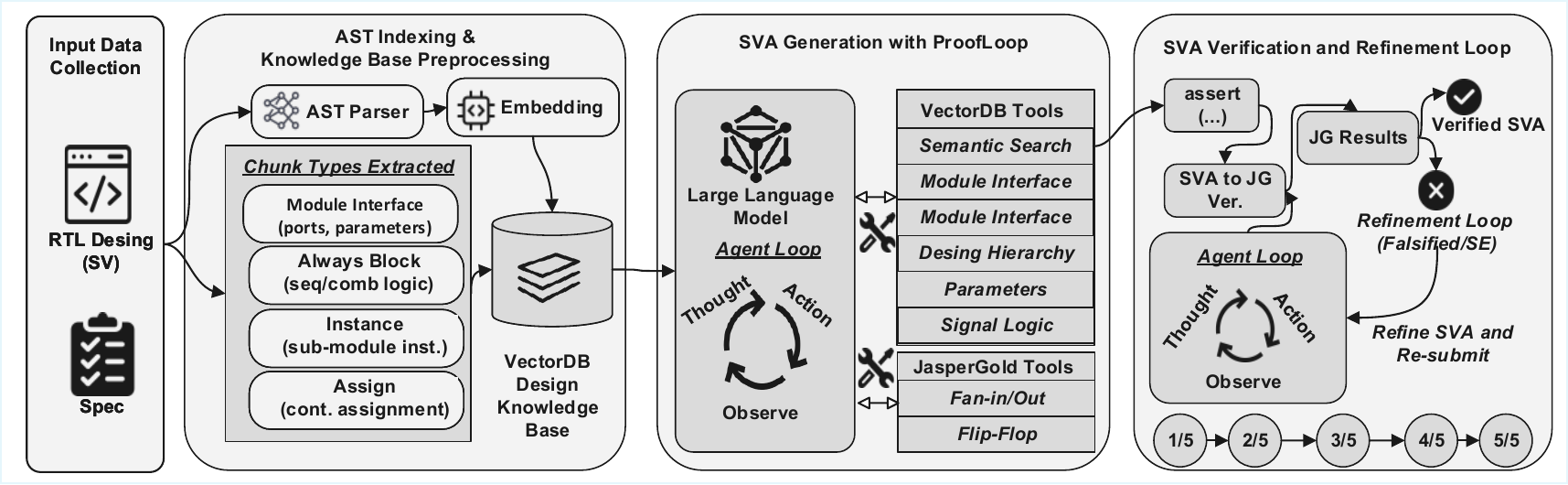}
\caption{\texttt{\textbf{ProofLoop}} framework (Tool-augmented ReAct agent). In Phase~A, the agent autonomously queries retrieval and structural analysis tools to gather design context. In Phase~B, the agent generates SVA assertions and iteratively refines them using JasperGold formal verification feedback.}
\label{fig:framework}
\end{figure*}

\begin{table}[t]
\scriptsize
\centering
\caption{Comparison of LLM-Based SVA Generation Approaches.}
\label{tab:related_comparison}
\setlength\tabcolsep{2.5pt}
\renewcommand{\arraystretch}{1.1}
\begin{tabular}{@{} l c c c c c @{}}
\toprule
\textbf{Method} & \textbf{RAG} & \textbf{JG Meta.} & \textbf{Agentic Tools} & \textbf{Solver Loop} & \textbf{Bench. Eval.} \\
\cmidrule(r){1-1} \cmidrule(r){2-2} \cmidrule(r){3-3} \cmidrule(r){4-4} \cmidrule(r){5-5} \cmidrule(r){6-6}
ChIRAAG \cite{mali2024chiraag}          & \ding{55} & \ding{55} & \ding{55} & Sim.  & \ding{55} \\
AssertLLM \cite{fang2024assertllm}      & \ding{55} & \ding{55} & \ding{55} & \ding{55}    & \ding{55} \\
SANGAM \cite{gupta2025sangam}           & \ding{51} & \ding{55} & \ding{55} & JG    & \ding{55} \\
Saarthi \cite{kumar2025saarthi}         & \ding{55} & \ding{55} & \ding{51} & FV    & \ding{55} \\
AssertionForge \cite{bai2025assertionforge} & KG & \ding{55} & \ding{55} & \ding{55} & \ding{55} \\
SVAgent \cite{guo2025svagent}           & \ding{55} & \ding{55} & \ding{55} & \ding{55} & \ding{55} \\
ChatSVA \cite{chatsva2026}             & \ding{51} & \ding{55} & \ding{55} & \ding{55} & \ding{55} \\ \cmidrule(r){1-1} \cmidrule(r){2-2} \cmidrule(r){3-3} \cmidrule(r){4-4} \cmidrule(r){5-5} \cmidrule(r){6-6}
\texttt{\textbf{ProofLoop}} (\textbf{Ours})                            & \ding{51} & \ding{51} & \ding{51} & JG    & \ding{51} \\
\bottomrule
\end{tabular}
\end{table}

\section{Proposed Method: \texttt{\textbf{ProofLoop}}}

\subsection{Overview}

Fig.~\ref{fig:framework} shows the \texttt{\textbf{ProofLoop}} framework. Given an RTL design and a natural-language assertion specification, the system operates in 3 stages: (1)~AST indexing builds a semantic knowledge base from the RTL, (2)~a ReAct agent gathers design context via tool calls in \textit{Phase~A}, and (3)~the agent generates SVA, followed by iteratively refining it using JasperGold formal verification in \textit{Phase~B} (see Alg. \ref{alg:react_agent}).

\subsection{AST Indexing and Knowledge Base}

The RTL design is parsed using a SystemVerilog compiler \cite{wolf2013yosys} to produce an verifiable AST. The AST is decomposed into four semantic chunk types:
\begin{enumerate}[leftmargin=*, nosep]
    \item \textbf{Module Interface}: ports, parameters, and localparams.
    \item \textbf{Always Block}: capturing sequential/combinational logic.
    \item \textbf{Instance}: sub-module instantiations with port connections.
    \item \textbf{Assign}: continuous assignment groups.
\end{enumerate}
Each chunk is embedded using an embedding model and stored in a vector database, creating a per-design knowledge base that supports semantic similarity search. This modular decomposition ensures that the LLM context window is populated only with the most relevant design fragments rather than the entire RTL, which is critical for multi-module designs where the full source can exceed context limits.
\subsection{Phase A: Context Gathering}

The tool autonomously gathers design context through an interleaved \textit{Thought} $\to$ \textit{Action} $\to$ \textit{Observation} loop for up to 6 rounds. The system has access to two categories of tools:

\noindent \textbf{(1) Retrieval tools} provide lightweight access to the vector database. They enable semantic search over the AST-indexed chunks, module interface queries (ports, parameters), hierarchy traversal across the dependency tree, parameter/localparam resolution, and signal-specific always-block retrieval. 

\textbf{Structural analysis tools} interface with JasperGold to extract hardware-level metadata that cannot be inferred from RTL text alone. These include signal fan-in and fan-out cone extraction, which reveals datapath dependencies, and flip-flop property inspection, which reports clock domains, reset behavior, and data-input connections. The system autonomously decides which tools to call based on the specification. This approach skips extra queries for simple cases while allowing deeper exploration for complex designs.

\subsection{Phase B: SVA Generation and Verification Loop}

After context gathering, the agent generates candidate SVA conditioned on the retrieved design context and the natural-language specification. The generated assertions are embedded into the RTL and passed to JasperGold for bounded formal verification. The solver attempts to prove all properties within a bounded time limit, classifying each as \textit{proven}, \textit{falsified}, or \textit{undetermined}.  For \textit{proven} or \textit{undetermined} properties, we perform an additional vacuity analysis to ensure that the proof is non-trivial (i.e., not satisfied due to unreachable antecedents or inactive triggering conditions). Otherwise, the solver's error messages and counterexample traces are fed back to the agent, enabling targeted repair of the failing assertions rather than blind regeneration. This feedback is programmatically parsed and incorporated into the agent’s context for targeted refinement. Rather than regenerating assertions from scratch, the agent performs localized edits, e.g., correcting signal references, aligning clock/reset domains, refining temporal operators, or restructuring antecedent/consequent conditions to match observed behavior. This solver-in-the-loop is a key differentiator from prior work: the system receives concrete, machine-generated feedback about \textit{why} an assertion fails, syntax errors, signal mismatches, or temporal violations, and can surgically correct the issue.

\begin{algorithm}[t]
\scriptsize
\caption{Tool-in-the-Loop Calls in \texttt{\textbf{ProofLoop}}.}
\label{alg:react_agent}
\begin{algorithmic}[1]
\REQUIRE RTL design $D$, NL specification $s$, LLM $\mathcal{L}$, tools $\mathcal{T}$
\ENSURE Verified SVA assertion

\STATE \textbf{AST Indexing:}
\STATE $\text{chunks} \gets \text{pyslang\_parse}(D)$ \COMMENT{4 chunk types}
\STATE $\text{VectorDB}.\text{store}(\text{embed}(\text{chunks}))$

\STATE \textbf{Phase A: Context Gathering} (up to 6 rounds)
\STATE $\text{conv} \gets [\text{system\_prompt}, \text{spec}(s), \text{design\_context}(D)]$
\FOR{$r = 1$ to $6$}
    \STATE $\text{msg} \gets \mathcal{L}(\text{conv}, \text{tools}=\mathcal{T})$ \COMMENT{ReAct: Thought+Action}
    \IF{no tool calls in msg}
        \STATE \textbf{break} \COMMENT{Agent has enough context}
    \ENDIF
    \FOR{each tool call $t$ in msg}
        \STATE $\text{obs} \gets \mathcal{T}.\text{execute}(t)$ \COMMENT{Observation}
        \STATE $\text{conv}.\text{append}(\text{obs})$
    \ENDFOR
\ENDFOR

\STATE \textbf{Phase B: Generation + Verification} (up to 3 rounds)
\STATE $\text{sva} \gets \mathcal{L}.\text{generate\_sva}(\text{conv})$
\FOR{$v = 1$ to $3$}
    \STATE $\text{result} \gets \text{JG}.\text{prove}(\text{sva}, D)$
    \IF{result.status $\in$ \{proven, undetermined\}}
        \STATE $\text{vacuity} \gets \text{JG}.\text{vacuity\_check}(\text{sva}, D)$
        \STATE \textbf{break}
    \ENDIF
    \STATE $\text{sva} \gets \mathcal{L}.\text{fix}(\text{sva}, \text{result.feedback})$
\ENDFOR
\RETURN sva, result
\end{algorithmic}
\end{algorithm}

\section{Experimental Results}

\subsection{Experimental Setup}

We evaluate \texttt{\textbf{ProofLoop}} on the FVEval benchmark \cite{kang2024fveval}, which comprises 96 pipeline and 96 FSM designs of varying complexity. We measure and report two primary metrics: (i) \textbf{Syntax}, where 1.0 denotes if JasperGold reports no compilation errors; and (ii) \textbf{Functionality}, which determines the ratio of proven to total number of assertions. Our primary model is Qwen3.5-35B with native tool-calling capabilities. To assess generality, we additionally evaluate Mistral-Small-3.1-24B under the same pipeline configuration. The baseline corresponds to direct prompting, where the LLM is provided with raw RTL and the natural-language specification without access to retrieval, structural tools, or verification feedback. All experiments are conducted on NVIDIA H100 GPUs.
\subsection{Assertion Verification results}

Table~\ref{tab:main_results} presents the main comparison across 192 designs. The tool-augmented pipeline substantially outperforms the direct prompting baseline in both syntax and functional correctness, nearly doubling the functionality score with Qwen3.5-35B. A cross-model evaluation with Mistral confirms that the improvement generalizes: the pipeline yields a comparable relative gain despite the smaller model's lower capability.

\begin{table}[H]
\scriptsize
\centering
\caption{Assertion Verification results: Pipeline vs.\ Baseline}
\label{tab:main_results}
\setlength\tabcolsep{6pt}
\renewcommand{\arraystretch}{1.15}
\begin{tabular}{@{} l l c c c c @{}}
\toprule
\textbf{Model} & \textbf{Method} & \textbf{Syntax} & \textbf{Functionality} & \textbf{Proven} & \textbf{Falsified} \\
\cmidrule(r){1-1} \cmidrule(r){2-2} \cmidrule(r){3-3} \cmidrule(r){4-4} \cmidrule(r){5-5} \cmidrule(r){6-6}
Qwen3.5-35B & Baseline & 78.3 & 43.2 & 445 & 246 \\
Qwen3.5-35B & Pipeline & \textbf{93.7} & \textbf{82.0} & \textbf{790} & 260 \\
\cmidrule(r){1-1} \cmidrule(r){2-2} \cmidrule(r){3-3} \cmidrule(r){4-4} \cmidrule(r){5-5} \cmidrule(r){6-6}
Mistral-Small-24B & Baseline & 65.8 & 31.1 & 249 & 200 \\
Mistral-Small-24B & Pipeline & 85.5 & 53.3 & 419 & 724 \\
\bottomrule
\end{tabular}
\end{table}

\subsection{Component-Wise Ablation Analysis}

To isolate each component's contribution, we conduct an ablation study on a 36-design subset, progressively removing pipeline components. Table~\ref{tab:ablation} presents the results.

\begin{table}[b]
\scriptsize
\centering
\caption{Ablation Study for Tool Augmentation and Solver Feedback.}
\label{tab:ablation}
\setlength\tabcolsep{6pt}
\renewcommand{\arraystretch}{1.15}
\begin{tabular}{@{} l c c c @{}}
\toprule
\textbf{Configuration} & \textbf{Syntax (\%)} & \textbf{Functionality (\%)} & \textbf{$\Delta$ vs.\ Full} \\
\cmidrule(r){1-1} \cmidrule(r){2-2} \cmidrule(r){3-3} \cmidrule(r){4-4}
Full Pipeline & \textbf{93.7} & \textbf{82.0} & --- \\
\quad $-$ RAG retrieval & 90.9 & 70.9 & $-$13.5\% \\
\quad $-$ Verification loop & 82.9 & 64.7 & $-$21.1\% \\
\quad $-$ Structural analysis tools & 80.0 & 58.4 & $-$28.8\% \\
\quad $-$ All tools (baseline) & 70.3 & 36.1 & $-$56.0\% \\
\bottomrule
\end{tabular}
\end{table}

Stripping away all tools reduces the pipeline to baseline-level performance, confirming that the agentic tool augmentation is the primary driver of improvement. Among individual components, the iterative verification loop provides the largest single contribution by enabling the agent to correct syntax errors and temporal mismatches through solver feedback. Structural analysis tools offer the next-largest gain, as they supply authoritative signal dependency and register information that the LLM cannot infer from RTL text alone. RAG retrieval provides a more modest but consistent benefit, primarily helping the agent resolve signal names and module interfaces accurately.

\subsection{Design Complexity Analysis}

\begin{figure}[t]
\centering
\includegraphics[width=\columnwidth]{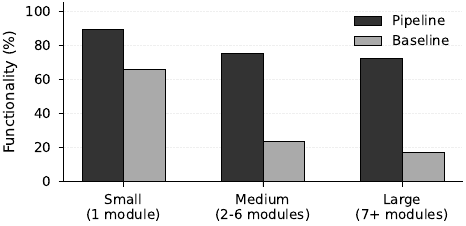}
\caption{Functionality by design complexity. The pipeline's advantage grows substantially with design size, as the baseline struggles with multi-module designs that exceed LLM context capacity.}
\label{fig:complexity}
\end{figure}

Fig.~\ref{fig:complexity} reveals that the pipeline's advantage scales with design complexity. For single-module designs, the baseline achieves 66.1\% functionality since the full RTL fits within the LLM's context window. However, as designs grow to include multiple interconnected modules, the baseline's performance degrades for large designs while the pipeline maintains relatively stable accuracy. This shows that AST-based retrieval and structural analysis are most effective for complex, multi-module designs. The pipeline's modular decomposition ensures bounded context per query, regardless of total design size.

\subsection{Design2SVA Benchmark Comparison}

We evaluated the \texttt{\textbf{ProofLoop}} framework on the Design2SVA task from FVEval \cite{kang2024fveval}. For each design, we perform 5 independent trials per design with nucleus sampling. Table~\ref{tab:fveval} compares against the FVEval baselines.

\begin{table}[b]
\scriptsize
\centering
\caption{FVEval Design2SVA Comparison.}
\label{tab:fveval}
\setlength\tabcolsep{11pt}
\begin{tabular}{@{} l cc cc @{}}
\toprule
 & \multicolumn{2}{c}{\textbf{Pipeline Designs}} & \multicolumn{2}{c}{\textbf{FSM Designs}} \\
\cmidrule(lr){2-3} \cmidrule(lr){4-5}
\textbf{Method} & \textbf{Func@1} & \textbf{Func@5} & \textbf{Func@1} & \textbf{Func@5} \\
\cmidrule(r){1-1} \cmidrule(r){2-2} \cmidrule(r){3-3} \cmidrule(r){4-4} \cmidrule(r){5-5}
GPT-4o \cite{kang2024fveval}           & 0.104 & 0.427 & 0.373 & 0.900 \\
Gemini-1.5-Pro \cite{kang2024fveval}   & 0.175 & 0.500 & 0.427 & 0.906 \\
Gemini-1.5-Flash \cite{kang2024fveval} & 0.025 & 0.125 & 0.079 & 0.281 \\
Mixtral-8x22B \cite{kang2024fveval}    & 0.119 & 0.472 & 0.054 & 0.167 \\
Llama-3.1-70B \cite{kang2024fveval}    & 0.167 & 0.615 & 0.231 & 0.719 \\
Llama-3.1-8B \cite{kang2024fveval}     & 0.150 & 0.552 & 0.121 & 0.521 \\
\midrule
\textbf{Ours (Qwen3.5-35B)}            & \textbf{0.412} & \textbf{0.896} & \textbf{0.860} & \textbf{0.958} \\
\cmidrule(r){1-1} \cmidrule(r){2-2} \cmidrule(r){3-3} \cmidrule(r){4-4} \cmidrule(r){5-5}
\end{tabular}
\end{table}

Despite using a smaller model, our pipeline substantially outperforms in both design categories. The improvement is particularly pronounced on pipeline designs, where multi-module structural complexity makes tool-augmented context gathering essential; the FVEval baselines rely on direct prompting and lack access to structural metadata or iterative solver feedback. The strong pass@5 scores indicate that with multiple trials, the agent reliably produces at least one formally provable assertion, suggesting robust performance even when individual attempts may vary. This shows that the combination of structured context and solver-guided refinement constrains the search space, enabling convergence to formally provable assertions even when individual generations may vary.

\subsection{Verification Loop Analysis}

\begin{figure}[t]
\centering
\includegraphics[width=\columnwidth]{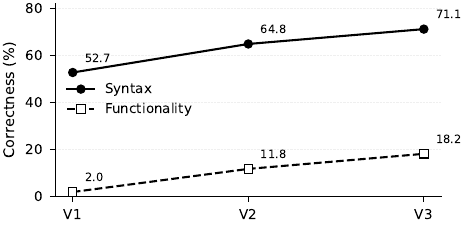}
\caption{Syntax and functional correctness improvement across verification rounds for the multi-iteration specs. Each round of solver feedback progressively improves both metrics, with syntax recovery.}
\label{fig:iterations}
\end{figure}

Fig.~\ref{fig:iterations} analyzes the iterative verification loop across all specifications. Two-thirds of specs (384) succeed on the first verification attempt. Of the remaining specs converge after two rounds and 114 require all three rounds. The agent is most effective at recovering from syntax and compilation errors: 59.3\% of syntax failures are corrected through solver feedback. Falsification recovery is harder; only 20 of 101 specs with initially falsified properties fully recover. This asymmetry suggests that LLMs are more capable of fixing structural mistakes given explicit error messages than of correcting temporal logic errors, which require deeper reasoning about hardware behavior.

\subsection{Case Study}

We illustrate the pipeline on a 6-module FSM design with a top-level module instantiating five execution units. The specification requires verifying that all pipeline stage registers clear correctly on reset: synchronous register must clear on the next clock edge, while asynchronous registers must clear immediately. The baseline produces syntactically invalid SVA because it cannot determine which registers use synchronous vs.\ asynchronous reset from the raw RTL alone.

In Phase~A, the framework issues 22 tool calls across 6 reasoning rounds: it queries module interfaces and parameters for all five units, performs semantic searches for register behavior and reset semantics, and retrieves specific always-block code to confirm clocking. In Phase~B, the first SVA attempt uses array-slice notation (\texttt{ready[1:3]\,==\,'d0}), which JasperGold rejects as a compilation error. The solver feedback identifies the invalid syntax, and the agent corrects it by expanding to individual element comparisons (\texttt{ready[1]\,==\,'d0\,\&\&\,ready[2]\,==\,'d0\,\&\&\,...}). The corrected assertions are formally proven in the second round (5 proven, 0 falsified). This example highlights how the tool-augmented context gathering resolves cross-module structural questions and the verification loop catches and corrects SVA syntax errors that would otherwise persist.

\section{Conclusion}
This paper presented a tool-augmented framework for automated SVA generation that integrates JasperGold as a solver-in-the-loop. The agent autonomously gathers design context through a suite of retrieval and structural analysis tools, then generates and iteratively refines SVA using formal proof feedback. The pipeline achieves 93.7\% syntax and 82.0\% functional correctness, outperforming the baseline on 66.1\% of designs with gains that scale with design complexity reaching significant relative improvement on large multi-module designs. On the Design2SVA benchmark, the pipeline achieves a combined pass@1 of 0.636, substantially outperforming baselines. The results suggest that the agentic tool-augmented approach addresses the fundamental limitation of prompt-only methods: the inability to access the structural metadata that verification engineers routinely rely on. Future work will explore domain-specific fine-tuning on curated SVA datasets, extension to SoC-scale multi-file designs, integration with open-source  formal engines to broaden accessibility, and investigate multi-agent collaboration, where agents handle distinct verification subtasks, enabling more comprehensive assertion suites for industrial-scale designs.

\bibliographystyle{IEEEtran}
\bibliography{refs}

@inproceedings{mali2024chiraag,
  title={{ChIRAAG}: {ChatGPT} Informed Rapid and Automated Assertion Generation},
  author={Mali, Bhabesh and Maddala, Karthik and Gupta, Vatsal and Reddy, Sweeya and Karfa, Chandan and Karri, Ramesh},
  booktitle={2024 IEEE Computer Society Annual Symposium on VLSI (ISVLSI)},
  pages={680--683},
  year={2024},
  organization={IEEE}
}

@article{fang2024assertllm,
  title={{AssertLLM}: Generating and Evaluating Hardware Verification Assertions from Design Specifications via Multi-{LLMs}},
  author={Fang, Wenji and Li, Mengming and Li, Min and Yan, Zhiyuan and Liu, Shang and Zhang, Hongce and Xie, Zhiyao},
  journal={arXiv preprint arXiv:2402.00386},
  year={2024}
}

@article{gupta2025sangam,
  title={{SANGAM}: {SystemVerilog} Assertion Generation via {Monte Carlo} Tree Self-Refine},
  author={Gupta, Adarsh and Mali, Bhabesh and Karfa, Chandan},
  journal={arXiv preprint arXiv:2506.13983},
  year={2025}
}

@article{kumar2025saarthi,
  title={Saarthi: The First {AI} Formal Verification Engineer},
  author={Kumar, Aman and Gadde, Deepak Narayan and Radhakrishna, Keerthan Kopparam and Lettnin, Djones},
  journal={arXiv preprint arXiv:2502.16662},
  year={2025}
}

@inproceedings{bai2025assertionforge,
  title={{AssertionForge}: Enhancing Formal Verification Assertion Generation with Structured Representation of Specifications and {RTL}},
  author={Bai, Yunsheng and Bany Hamad, Ghaith and Suhaib, Syed and Ren, Haoxing},
  booktitle={2025 IEEE International Conference on LLM-Aided Design (ICLAD)},
  year={2025},
  organization={IEEE}
}

@article{guo2025svagent,
  title={{SVAgent}: {AI} Agent for Hardware Security Verification Assertion},
  author={Guo, Rui and Ayalasomayajula, Avinash and Li, Henian and Zhou, Jingbo and Saha, Sujan Kumar and Farahmandi, Farimah},
  journal={arXiv preprint arXiv:2507.16203},
  year={2025}
}

@inproceedings{pulavarthi2025assertionbench,
  title={{AssertionBench}: A Benchmark to Evaluate Large-Language Models for Assertion Generation},
  author={Pulavarthi, Vaishnavi and Nandal, Deeksha and Dan, Soham and Pal, Debjit},
  booktitle={Findings of the Association for Computational Linguistics: NAACL 2025},
  year={2025}
}

@article{kande2024security,
  title={(Security) Assertions by Large Language Models},
  author={Kande, Rahul and Pearce, Hammond and Tan, Benjamin and Dolan-Gavitt, Brendan and Thakur, Shailja and Karri, Ramesh and Rajendran, Jeyavijayan},
  journal={IEEE Transactions on Information Forensics and Security},
  year={2024},
  publisher={IEEE}
}

@article{orenes2023using,
  title={Using {LLMs} to Facilitate Formal Verification of {RTL}},
  author={Orenes-Vera, Marcelo and Martonosi, Margaret and Wentzlaff, David},
  journal={arXiv preprint arXiv:2309.09437},
  year={2023}
}

@article{menon2025vert,
  title={Enhancing Large Language Models for Hardware Verification: A Novel {SystemVerilog} Assertion Dataset},
  author={Menon, Aman and others},
  journal={ACM Transactions on Design Automation of Electronic Systems},
  year={2025},
  publisher={ACM}
}

@article{kang2024fveval,
  title={{FVEval}: Understanding Language Model Capabilities in Formal Verification of Digital Hardware},
  author={Kang, Minwoo and Liu, Mingjie and Bany Hamad, Ghaith and Suhaib, Syed and Ren, Haoxing},
  journal={arXiv preprint arXiv:2410.23299},
  year={2024}
}

@article{witharana2022survey,
  title={A Survey on Assertion-Based Hardware Verification},
  author={Witharana, Hasini and Lyu, Yangdi and Charles, Subodha and Mishra, Prabhat},
  journal={ACM Computing Surveys},
  volume={54},
  number={11s},
  pages={1--33},
  year={2022},
  publisher={ACM}
}

@inproceedings{aftabjahani2021special,
  title={Special Session: {CAD} for Hardware Security---Automation is Key to Adoption of Solutions},
  author={Aftabjahani, Sohrab and Kastner, Ryan and Tehranipoor, Mark and Farahmandi, Farimah and Oberg, Jason and Nordstrom, Anders and Fern, Nicole and Althoff, Alric},
  booktitle={2021 IEEE 39th VLSI Test Symposium (VTS)},
  pages={1--10},
  year={2021},
  organization={IEEE}
}

@inproceedings{farzana2019soc,
  title={{SoC} Security Verification Using Property Checking},
  author={Farzana, Nusrat and Rahman, Fahim and Tehranipoor, Mark and Farahmandi, Farimah},
  booktitle={2019 IEEE International Test Conference (ITC)},
  pages={1--10},
  year={2019},
  organization={IEEE}
}

@inproceedings{lewis2020retrieval,
  title={Retrieval-Augmented Generation for Knowledge-Intensive {NLP} Tasks},
  author={Lewis, Patrick and Perez, Ethan and Piktus, Aleksandara and Petroni, Fabio and Karpukhin, Vladimir and Goyal, Naman and K{\"u}ttler, Heinrich and Lewis, Mike and Yih, Wen-tau and Rockt{\"a}schel, Tim and others},
  booktitle={Advances in Neural Information Processing Systems (NeurIPS)},
  year={2020}
}

@inproceedings{wolf2013yosys,
  title={Yosys---A Free {Verilog} Synthesis Suite},
  author={Wolf, Clifford and Glaser, Johann and Kepler, Johannes},
  booktitle={Proceedings of the 21st Austrian Workshop on Microelectronics (Austrochip)},
  volume={97},
  year={2013}
}

@article{chatsva2026,
  title={{ChatSVA}: Bridging {SVA} Generation for Hardware Verification via Task-Specific {LLMs}},
  author={Fu, Lik Tung and Zhou, Jie and Ren, Shaokai and Zhang, Mengli and Xiong, Jia and Jiang, Hugo and Guan, Nan and Wang, Xi and Yang, Jun},
  journal={arXiv preprint arXiv:2604.02811},
  year={2026}
}

@article{deepv2025,
  title={{DeepV}: A Model-Agnostic Retrieval-Augmented Framework for {Verilog} Code Generation with a High-Quality Knowledge Base},
  author={Ibnat, Zahin and Calzada, Paul E. and Ihtemam, Rasin Mohammed and Saha, Sujan Kumar and Zhou, Jingbo and Farahmandi, Farimah and Tehranipoor, Mark},
  journal={arXiv preprint arXiv:2510.05327},
  year={2025}
}

@inproceedings{vasudevan2010goldmine,
  title={{GoldMine}: Automatic Assertion Generation Using Data Mining and Static Analysis},
  author={Vasudevan, Shobha and Sheridan, David and Patel, Sanjay and Tcheng, David and Tuohy, Bill and Johnson, Daniel},
  booktitle={2010 Design, Automation \& Test in Europe Conference \& Exhibition (DATE)},
  pages={626--629},
  year={2010},
  organization={IEEE}
}

@article{germiniani2022harm,
  title={{HARM}: A Hint-Based Assertion Miner},
  author={Germiniani, Samuele and Pravadelli, Graziano},
  journal={IEEE Transactions on Computer-Aided Design of Integrated Circuits and Systems},
  volume={41},
  number={11},
  pages={4277--4288},
  year={2022},
  publisher={IEEE}
}

@inproceedings{danese2017ateam,
  title={{A-Team}: Automatic Template-Based Assertion Miner},
  author={Danese, Alessandro and Dalla Riva, Nicol{\`o} and Pravadelli, Graziano},
  booktitle={Proceedings of the 54th Annual Design Automation Conference (DAC)},
  pages={1--6},
  year={2017},
  organization={ACM}
}

@inproceedings{orenes2021autosva,
  title={{AutoSVA}: Democratizing Formal Verification of {RTL} Module Interactions},
  author={Orenes-Vera, Marcelo and Manocha, Aninda and Wentzlaff, David and Martonosi, Margaret},
  booktitle={Proceedings of the 58th ACM/IEEE Design Automation Conference (DAC)},
  pages={535--540},
  year={2021},
  organization={ACM/IEEE}
}

@article{thakur2023verigen,
  title={{VeriGen}: A Large Language Model for {Verilog} Code Generation},
  author={Thakur, Shailja and Ahmad, Baleegh and Pearce, Hammond and Tan, Benjamin and Dolan-Gavitt, Brendan and Karri, Ramesh and Garg, Siddharth},
  journal={arXiv preprint arXiv:2308.00708},
  year={2023}
}

@article{tsai2023rtlfixer,
  title={{RTLFixer}: Automatically Fixing {RTL} Syntax Errors with Large Language Models},
  author={Tsai, Yun-Da and Liu, Mingjie and Ren, Haoxing},
  journal={arXiv preprint arXiv:2311.16543},
  year={2023}
}

@article{blocklove2025improving,
  title={Automatically Improving {LLM}-Based {Verilog} Generation Using {EDA} Tool Feedback},
  author={Blocklove, Jason and Thakur, Shailja and Tan, Benjamin and Pearce, Hammond and Garg, Siddharth and Karri, Ramesh},
  journal={ACM Transactions on Design Automation of Electronic Systems},
  year={2025},
  publisher={ACM}
}

@inproceedings{sun2023nl2sva,
  title={Towards Improving Verification Productivity with Circuit-Aware Translation of Natural Language to {SystemVerilog} Assertions},
  author={Sun, Chuyue and Hahn, Christopher and Trippel, Caroline},
  booktitle={Proceedings of the DAV Workshop},
  year={2023}
}

@inproceedings{kang2024domain,
  title={Domain-Adapted {LLMs} for {VLSI} Design and Verification: A Case Study on Formal Verification},
  author={Kang, Minwoo and Liu, Mingjie and Bany Hamad, Ghaith and Suhaib, Syed and Ren, Haoxing},
  booktitle={2024 IEEE 42nd VLSI Test Symposium (VTS)},
  year={2024},
  organization={IEEE}
}

@inproceedings{hassan2024llmguided,
  title={{LLM}-Guided Formal Verification Coupled with Mutation Testing},
  author={Hassan, Muhammad and Ahmadi-Pour, Sallar and Qayyum, Khushboo and Jha, Chandan Kumar and Drechsler, Rolf},
  booktitle={2024 Design, Automation \& Test in Europe Conference \& Exhibition (DATE)},
  pages={1--6},
  year={2024},
  organization={IEEE}
}

@book{clarke2018model,
  title={Handbook of Model Checking},
  author={Clarke, Edmund M and Henzinger, Thomas A and Veith, Helmut and Bloem, Roderick},
  year={2018},
  publisher={Springer}
}

@book{vijayaraghavan2005sva,
  title={A Practical Guide for {SystemVerilog} Assertions},
  author={Vijayaraghavan, Srikanth and Ramanathan, Meyyappan},
  year={2005},
  publisher={Springer}
}

@article{mashnoor2026meltrtl,
  title={MeltRTL: Multi-Expert LLMs with Inference-time Intervention for RTL Code Generation},
  author={Mashnoor, Nowfel and Akyash, Mohammad and Kamali, Hadi and Azar, Kimia},
  journal={arXiv preprint arXiv:2601.13015},
  year={2026}
}

@inproceedings{akyash2025rtl++,
  title={Rtl++: Graph-enhanced llm for rtl code generation},
  author={Akyash, Mohammad and Azar, Kimia and Kamali, Hadi},
  booktitle={2025 IEEE International Conference on LLM-Aided Design (ICLAD)},
  pages={44--50},
  year={2025},
  organization={IEEE}
}

@inproceedings{akyash2025decortl,
  title={Decortl: A run-time decoding framework for rtl code generation with llms},
  author={Akyash, Mohammad and Azar, Kimia and Kamali, Hadi},
  booktitle={2025 IEEE/ACM International Conference On Computer Aided Design (ICCAD)},
  pages={1--9},
  year={2025},
  organization={IEEE}
}

@article{liu2024rtlcoder,
  title={Rtlcoder: Fully open-source and efficient llm-assisted rtl code generation technique},
  author={Liu, Shang and Fang, Wenji and Lu, Yao and Wang, Jing and Zhang, Qijun and Zhang, Hongce and Xie, Zhiyao},
  journal={IEEE Transactions on Computer-Aided Design of Integrated Circuits and Systems},
  volume={44},
  number={4},
  pages={1448--1461},
  year={2024},
  publisher={IEEE}
}

@article{zhao2025codev,
  title={Codev: Empowering llms with hdl generation through multi-level summarization},
  author={Zhao, Yang and Huang, Di and Li, Chongxiao and Jin, Pengwei and Song, Muxin and Xu, Yinan and Nan, Ziyuan and Gao, Mingju and Ma, Tianyun and Qi, Lei and others},
  journal={IEEE Transactions on Computer-Aided Design of Integrated Circuits and Systems},
  year={2025},
  publisher={IEEE}
}

@inproceedings{mashnoor2025llm,
  title={Llm-ift: Llm-powered information flow tracking for secure hardware},
  author={Mashnoor, Nowfel and Akyash, Mohammad and Kamali, Hadi and Azar, Kimia},
  booktitle={2025 IEEE 43rd VLSI Test Symposium (VTS)},
  pages={1--5},
  year={2025},
  organization={IEEE}
}

@article{huang2024towards,
  title={Towards llm-powered verilog rtl assistant: Self-verification and self-correction},
  author={Huang, Hanxian and Lin, Zhenghan and Wang, Zixuan and Chen, Xin and Ding, Ke and Zhao, Jishen},
  journal={arXiv preprint arXiv:2406.00115},
  year={2024}
}

\end{document}